%% file: main.tex
\documentclass[a4paper,twoside]{article}

\usepackage{mathptmx}
\usepackage{calc}
\usepackage{graphicx}
\usepackage{amsmath}
\usepackage{amssymb}
\usepackage{amsthm}
\usepackage{booktabs}
\usepackage{tabularx}
\usepackage{enumitem}
\usepackage{subcaption}
\usepackage{algorithm2e}
\usepackage{xurl}
\usepackage{natbib}

\usepackage{multicol} 
\usepackage{colortbl} 

\usepackage{SCITEPRESS}

\renewcommand{\orcidAuthor}[1]{\includegraphics[width=0.3cm]{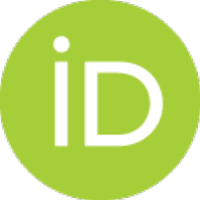}\thanks{\href{https://orcid.org/#1}{https://orcid.org/#1}}}

\usepackage{hyperref}

\begin{document}

\title{Exploring LLM Capabilities in Extracting DCAT-Compatible Metadata for Data Cataloging}

\author{\authorname{Lennart Busch\sup{1}\orcidAuthor{0009-0001-8952-3523}, Daniel Tebernum\sup{2}\orcidAuthor{0000-0002-4772-9099} and Gissel Velarde\sup{1}\orcidAuthor{0000-0001-5392-9540}}
\affiliation{\sup{1}IU International University of Applied Sciences, Erfurt, 99084, Germany}
\affiliation{\sup{2}Fraunhofer Institute for Software and Systems Engineering ISST, Dortmund, 44147, Germany}
\email{lennart.busch@iu-study.org, daniel.tebernum@isst.fraunhofer.de, gissel.velarde@iu.org}
}


\abstract{
Efficient data exploration is crucial as data becomes increasingly important for accelerating processes, improving forecasts and developing new business models.
Data consumers often spend 25-98\% of their time searching for suitable data due to the exponential growth, heterogeneity and distribution of data.
Data catalogs can support and accelerate data exploration by using metadata to answer user queries.
However, as metadata creation and maintenance is often a manual process, it is time-consuming and requires expertise.
This study investigates whether LLMs can automate metadata maintenance of text-based data and generate high-quality DCAT-compatible metadata.
We tested zero-shot and few-shot prompting strategies with LLMs from different vendors for generating metadata such as titles and keywords, along with a fine-tuned model for classification.
Our results show that LLMs can generate metadata comparable to human-created content, particularly on tasks that require advanced semantic understanding.
Larger models outperformed smaller ones, and fine-tuning significantly improves classification accuracy, while few-shot prompting yields better results in most cases.
Although LLMs offer a faster and reliable way to create metadata, a successful application requires careful consideration of task-specific criteria and domain context.
}

\onecolumn \maketitle \normalsize \setcounter{footnote}{0} \vfill

\input{sections/01_introduction}
\input{sections/02_background}
\input{sections/03_methodology}
\input{sections/05_discussion}

\input{sections/06_contribution}
\input{sections/08_conclusion}
\newpage

\bibliographystyle{apalike}
{\small
\bibliography{literature}}



\end{document}

%% file: sections/01_introduction.tex
\section{Introduction}
\label{sec:introduction}
Text-based data and documents play a critical role for companies and organizations, forming the backbone of numerous business processes and serving as the primary repository for business data \citep{rowe_handling_2012}. However, these documents are often difficult to locate. Studies reveal that knowledge workers can spend 25\%-98\% of their work time searching for information contained in documents \citep{deng2017data}, and up to 83\% have had to recreate existing documents because they were unable to find them within their organization's network \citep{m-files_2019_2019}. In a study for the European Commission, the consulting firm PwC estimated that the cost of such data that does not comply with the FAIR principles costs the European economy alone 10.2 billion Euro per year \citep{pwc_cost-benefit_2018}.

An effective solution for managing data and addressing associated challenges is the implementation of data catalogs.
It is a system that facilitates the findability, accessibility, and organization of data. It serves as a centralized platform for semantically classifying and organizing data sources \citep{kotsis_data_2021}. An important aspect of a data catalog is the use of metadata. However, the creation of metadata, particularly in large quantities for data catalogs, is a time-consuming process that relies heavily on the skills of individual users \citep{mondal_metadata_2018}. This dependency can lead to situations where data catalogs degrade into inoperable ``data swamps'' due to poor metadata quality  \citep{eichler_enterprise-wide_2021}.

Therefore, the development of automated processes for generating metadata has been an enduring area of research for many decades \citep{jenkins_automatic_1999}. The development of Transformer-based neural networks by researchers at Google in 2017 led to significant advancements in the field of natural language processing \citep{vaswani_attention_2017}. These advanced neural networks exhibit exceptional proficiency in natural language processing tasks, ranging from text classification to abstract writing and problem-solving \citep{hasselaar_jobs_2023}. 

Building on these advances and addressing the identified challenges, this study poses the following research question: \textit{How effective are autoregressive LLMs at generating descriptive, DCAT-compatible metadata from text-based documents with human-level quality?}

Based on this research question, three hypotheses (H) have been formulated and will be tested in this study:

\textbf{H1.} Autoregressive LLMs can generate descriptive, DCAT-compatible metadata with accuracy comparable to human-curated metadata, with the quality of the output improving when transitioning from zero-shot to few-shot prompting.

\textbf{H2.} Fine-tuning autoregressive models specifically for classification tasks further enhances accuracy and outperforms few-shot prompting in this scenario.

\textbf{H3.} Larger autoregressive models, such as GPT-4o, outperform smaller models, like Llama 3.2 3B, in generating high-quality and consistent metadata.

Our \textbf{contribution} lies in creating a comprehensive guide that outlines best practices for optimizing LLMs in metadata generation. It addresses gaps in the existing literature by providing practical strategies for effective LLM use, eliminating the need for extensive frameworks or overly complex adjustments. \footnote{All utilized prompts, results, and code are available in the following GitHub repository: \url{https://github.com/l94779589/Exploring-LLM-Capabilities-in-Extracting-DCAT-Compatible-Metadata-for-Data-Cataloging}}

This paper is organized as follows: Section 2 offers an overview of metadata, data catalogs, DCAT, and LLMs, along with related work to provide context for our study. Section 3 outlines the methodology, detailing the datasets, metrics, and experimental design. Section 4 presents the results, followed by a discussion of their theoretical and practical implications in Section 5. Finally, Section 6 summarizes the key findings, acknowledges limitations, and suggests directions for future research.

%% file: sections/02_background.tex
\section{Background}
\label{sec:background}

\textbf{Metadata} can be described as ``data about data'' \citep{pomerantz_metadata_2015, horodyski_metadata_2022} or, more precisely, as ``[...] structured information that describes, explains, locates, or otherwise makes it easier to retrieve, use, or manage an information resource.'' \citep[p. 1]{niso2004understanding}.
Metadata provides information about the data itself without having to process the actual data \citep{sabot2022importance}.
High-quality metadata enhances productivity, compliance, and scalability within organizations.
It e.g. facilitates rapid data retrieval and clarifies the applicability of legal regulations \citep{roszkiewicz2010enterprise}.
However, ``[m]etadata collection is expensive so incremental collection along the workflow is required'' \citep[p. 128]{jeffery2020}.
This, in conjunction with the need for high-quality metadata to achieve the described improvements \citep{quimbertJM0Z20}, is driving the further development of automation solutions in companies \citep{ochoaD09}.
Metadata can be categorized into different classes such as descriptive, administrative and structural metadata \citep{riley2017understanding}.
In this paper, we will focus on descriptive metadata, such as the title, description, and theme of an information object.

\textbf{Data Catalogs} ``collect, create and maintain metadata'' \citep[p. 141]{quimbertJM0Z20}.
By doing so, they streamline data discovery, access, and curation \citep{kotsis_data_2021, jahnke2023data} and thus support FAIR principles, \citep{labadie2020fair}, data governance \citep{shanmugam2016aspects} and data democratization \citep{Eichler202219}.
They establish the connection between data supply and data demand \citep{jahnke2023data}.
Data catalog solutions are highly diverse, with specialized implementations tailored to specific purposes \citep{zaidi2017data,jahnke2023data}.
So far, limited research has been conducted on the design and requirements of data catalogs.
However, the automation of metadata generation has shown to be an important feature \citep{petrikUB23,tebernum2024}.

\textbf{Data Catalog Vocabulary (DCAT)} is the de facto standard schema for metadata management in data catalogs \citep{albertoni2023w3c} and is managed by the World Wide Web Consortium \citep{world_wide_web_consortium_data_2024}.
It is built upon the resource-centric Dublin Core Metadata Initiative Metadata Terms (DCTERMS) and now emphasizes the description of datasets and data catalogs \citep{maali2010enabling}.
In this study, we will incorporate the following DCAT-Properties:
\begin{multicols}{2}
\begin{itemize}
    \item dcterms:title
    \item dcterms:description
    \item dcterms:creator
    \item dcterms:language
    \item dcterms:spatial
    \item dcat:issued
    \item dcat:keyword
    \item dcat:theme
\end{itemize}
\end{multicols}

\textbf{Large Language Models (LLMs)} are language models that employ transformer-based neural networks \citep{vaswani_attention_2017} and leverage deep learning algorithms \citep{hadi_large_2023}. These models have been pre-trained \citep{jurafsky_speech_2024} on large amounts of textual data  \citep{hadi_large_2023}, enabling them to generate text and perform diverse language-based tasks with exceptional, human-level performance \citep{hadi_large_2023}. In performing these tasks, they demonstrate strong thinking, planning, and decision-making skills, even if they have not been specifically trained for a particular task \citep{naveed_comprehensive_2024}.

Modern, commercial LLMs are predominantly decoder-only models, designed to generate text by predicting one token at a time based on preceding tokens in a sequence. This autoregressive architecture excels at generating fluent, contextually appropriate text and is highly adaptable across various downstream tasks \citep{wang_what_2022}. This study incorporates the following decoder-only models:

\begin{enumerate}
    \item \textbf{Llama 3.1 8B \& Llama 3.1 70B}, released by Meta in July 2024, supports a context window of 128,000 tokens \citep{meta_introducing_2024}.
    \item \textbf{Llama 3.2 3B}, released by Meta in September 2024, is optimized for mobile devices. This model supports a context window of up to 128,000 tokens and was created through knowledge distillation and structured pruning of the Llama 3.1 8B and 70B models \citep{meta_llama_2024}.
    \item \textbf{Google Gemini 1.5 Flash \& Pro}, introduced by Google in March 2024, consists of two variants: the smaller Flash and the larger Pro. Gemini 1.5 supports a context length of up to 2,000,000 tokens \citep{gemini_team_gemini_2024}.
    \item \textbf{GPT-4o \& GPT-4o Mini}, developed by OpenAI and released in July and August 2024, both support a context length of 128,000 tokens. \citep{openai_inc_gpt-4o_2024, openai_inc_openai_2024-1}.
\end{enumerate}

\textbf{Related work} on LLMs has increased exponentially, with the number of research papers growing from 42 in 2018 to over 28,000 in 2024.
However, work related to metadata and data catalogs remains fragmented and, in addition to new methods and frameworks, is focused on specific elements such as keywords and classification.

Despite this growth in LLM research, the field around metadata generation for data catalogs remains fragmented.
Existing studies focus on isolated elements such as keywords and classification.
While work such as PromptRank \citep{kong_promptrank_2023}, LaFiCMIL \citep{sun_laficmil_2024}, and efforts to improve FAIR-compliant ecosystems \citep{arnold_llm_fair_2024} represent valuable advances. However, these initiatives primarily address specific technical challenges, methods, or frameworks, and none of them adopt a truly comprehensive and practical approach that integrates multiple DCAT metadata properties, diverse LLM model families, and size variants, while ensuring broad accessibility to a wide range of users.

%% file: sections/03_methodology.tex
\section{Method}
\label{sec:methodology}

\begin{figure*}[ht]
\centering
\includegraphics[width=1.0\textwidth]{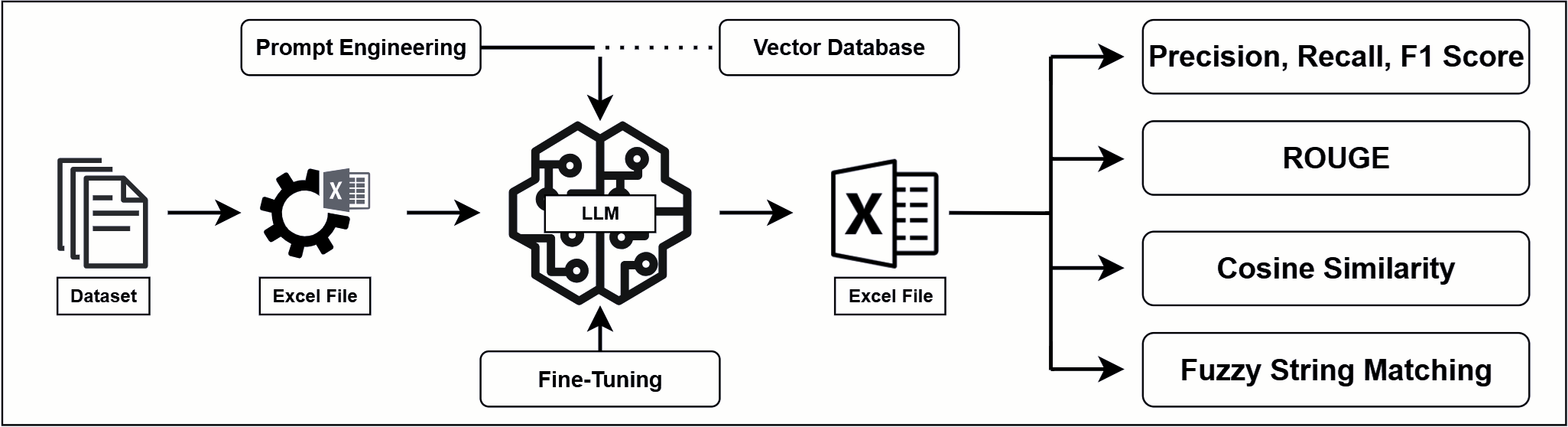} 
\caption{Method design.}
\label{fig:method}
\end{figure*}

The method design is illustrated in Figure \ref{fig:method}. 
We converted each dataset into an Excel file, processed each item in the dataset with an LLM, and stored the LLM's responses alongside the human-annotated ground truth in a separate Excel file for subsequent evaluation. For each DCAT property, at least one dataset was utilized, see Section 3.1. The same prompt design was applied across all LLMs for each DCAT property and dataset. For straightforward information extraction (e.g., dcterms:creator), we used predefined examples in few-shot prompts; for more complex, context-dependent properties, dynamic few-shot prompts were employed. Examples were retrieved from a vector database—which stores data as high-dimensional vectors for similarity searches—drawn either from the validation subset or a thematically similar dataset. The embedding model used was Google’s “text-embedding-004.” \citep{google_llc_text_embeddings_2024}. All LLMs used the same settings: for all DCAT properties, the temperature was set to 0 for reproducibility, and a repetition penalty of 0.5 was applied specifically for dcat:keyword to encourage diversity. The LLM’s role was chosen based on its task, reflecting role sensitivity in LLMs \citep{zheng_is_2023}.

For the dcat:theme property, a Gemini 1.5 Flash model was fine-tuned, as classification tasks require heightened domain-specific precision compared to other metadata properties \citep{chalkidis_large-scale_2019}.
Due to limited resources, we did not perform fine-tuning on other attributes such as abstracts and keywords, nor did we perform such procedures on additional models.

\subsection{Datasets}
We evaluated 9 \textbf{datasets} across all eight DCAT-properties, including \citep{wang_paper_2018, cohan_discourse-aware_2018, goyal_flores-101_2021, yucheng_arxiv_latest_2024, hattori_iges_2022, kim_semeval-2010_2010, augenstein_semeval_2017, chalkidis_large-scale_2019, singh_sector_data_2023}.

\textbf{dcterms:title}:
The dataset comprises 10,874 title-and-abstract pairs sourced from the ACL Anthology Network. On average, the titles and abstracts consist of 9 and 116 words each \citep{wang_paper_2018}. The dataset was utilized by \cite{mishra_automatic_2021}, whose results will serve as a baseline for this work. Few-shot prompting was implemented using a dynamic retrieval approach. The dataset for the examples consisted of 5,714 abstract-title pairs sourced from two other ACL Anthology datasets \citep{aclmeeting_acl2023-papers_2023, aclmeeting_acl2024-papers_2024}. 

\textbf{dcterms:description}:
The dataset includes 133,000 papers and abstracts from PubMed.com and is split into training (94\%), validation (3\%), and test (3\%) subsets \citep{cohan_discourse-aware_2018}. The papers and abstracts have an average length of 3,016 and 203 words, respectively. The dataset was also utilized by \cite{guo_longt5_2021}; their findings will serve as a benchmark for this research. For few-shot prompting, the validation subset of the dataset was employed. 

\textbf{dcterms:creator}:
The dataset consists of 1,160 scientific papers from ArXiv.com \citep{yucheng_arxiv_latest_2024}. It includes a total of 7,438 authors. Only the first page of each paper was presented to the LLMs. Few-shot prompting employed a predefined approach, where two examples were presented to the model. 

\textbf{dcterms:language}:
The dataset includes 3001 sentences sourced from English Wikipedia. Each sentence was translated into 101 languages by professional translators \citep{goyal_flores-101_2021}. For this study, a subset of 5 sentences with their full set of 101 translations was selected. Few-shot prompting used a dynamic retrieval approach. The dataset for the examples included two other example sentences along with their 101 translations from the original dataset. 

\textbf{dcterms:spatial}: 
The dataset initially contained 144 country-specific submissions to the UN \citep{hattori_iges_2022}. After excluding non-English documents and duplicate EU submissions, 102 examples remained. To expand the dataset, we generated two additional versions of each example by translating it from English to another language and back to English using the Google Translate API, increasing the total to 306 examples. Few-shot prompting employed a predefined approach, where two examples were presented to the model.

\textbf{dcat:issued}:
The dataset for this DCAT property is the same as the one used for ``dcterms:creator''. Few-shot prompting employed a predefined approach, where two examples were presented to the models. 

\textbf{dcat:keyword}:
Two datasets were utilized:

\begin{itemize}
    \item The SemEval2010 dataset consists of 243 full scientific papers. Each paper, approximately 8,332 tokens in length, is annotated with keywords provided by the authors and professional editors \citep{kim_semeval-2010_2010}.

  \item The SemEval2017 dataset consists of 493 paragraphs from ScienceDirect journal articles, averaging 178 tokens each. Keywords were annotated by both an undergraduate student and an expert \citep{augenstein_semeval_2017}.
\end{itemize}

Both datasets have been widely used in subsequent research, including the work of \cite{kong_promptrank_2023}, their findings will serve as a benchmark in this study.

Few-shot prompting was implemented using a dynamic retrieval approach, where three relevant examples were sourced from a third dataset called ``Inspec'', which comprises 2,000 abstracts of scientific journal papers \citep{hulth_improved_2003}.

\textbf{dcat:theme}:
Two datasets were utilized:

\begin{itemize}
    \item  EURLEX57K consists of 57,000 legal documents from the EU and is annotated using the EuroVoc thesaurus. While the EuroVoc thesaurus encompasses over 7,000 labels and their IDs, only 4,271 were actively assigned in this dataset. This dataset is divided into training (45,000 examples), validation (6,000 examples), and test (6,000 examples) subsets.  The dataset was curated and first utilized by \cite{chalkidis_large-scale_2019} and later reused by \cite{sun_laficmil_2024}, both results will serve as benchmarks for this study.
    
    \item The GIZ Sector Data dataset, comprises 22,674 items, divided into training (10,015 examples), validation (11,754 examples), and test (905 examples) subsets \citep{singh_sector_data_2023}. Each item in the test subset is annotated with labels from a set of 16 distinct labels, with an average text length of approximately 50 words per item.  
    This dataset was utilized by \cite{singh_mpnet-multilabel-sector-classifier_2023}, whose results will serve as a benchmark. 
\end{itemize}

Few-shot prompting used a dynamic retrieval approach, selecting three examples from the validation subset of each dataset. Due to the extensive number of labels in the EUROLEX57K dataset, the workflow applied here differed from the other DCAT properties and involved four steps for few-shot prompting: (1) The LLM generated up to 10 keywords/phrases from the input text; (2) each keyword was matched against a vector database, which stored the labels individually, to find the two most relevant labels; (3) three examples were retrieved from another vector database based on the current input; and (4) the model received the combined input text, retrieved examples, and labels for final classification.

Furthermore, fine-tuning was applied to Gemini 1.5 Flash in Google AI Studio for this classification task, using the respective training subset of the datasets with the following settings:

\begin{itemize}
    \item EURLEX57K: 5 epochs with a learning rate of 0.0006 and a batch size of 16. 
    \item GIZ Sector Data: 8 epochs with a learning rate of 0.0003 and a batch size of 16. 
\end{itemize}

\subsection{Evaluation Metrics}
Due to the heterogeneous nature of the DCAT properties included in this study, multiple evaluation metrics were employed. However, not all metrics were applicable to every property, as certain properties cannot be meaningfully assessed using specific metrics. The following section provides a detailed description of each metric and specifies the corresponding properties to which it was applied.

\textbf{Precision, Recall and F Score} are metrics for assessing binary classification performance. In multi-class classification, micro F1 extend these metrics \citep{grandini_metrics_2020}. Precision measures the proportion of correctly retrieved results among all retrieved instances, while recall evaluates the ability to identify all relevant instances \citep{dalianis_clinical_2018}. The F1 score balances precision and recall as their harmonic mean.  The micro F1 score aggregates true positives, false positives, and false negatives across all classes, reflecting overall performance with greater weight on larger classes \citep{grandini_metrics_2020}. These metrics were applied to dcterms:creator, dcterms:language, dcterms:spatial, dcat:issued, dcat:keyword and dcat:theme. This metric was selected to ensure consistency with comparison benchmarks employed in this study and due to its widespread adoption in the field of NLP research.

\begin{table*}[t]
    \centering
    \small 
    \setlength{\tabcolsep}{4pt} 

    \begin{tabular}{@{}llccccccccc@{}}
        \toprule
        \textbf{Model} & \textbf{Prompting} & \multicolumn{2}{c}{\textbf{ROUGE-1}} & \multicolumn{2}{c}{\textbf{ROUGE-2}} & \multicolumn{2}{c}{\textbf{ROUGE-L}} & \multicolumn{2}{c}{\textbf{Cosine Sim.}} \\ 
        \cmidrule(r){3-4} \cmidrule(r){5-6} \cmidrule(r){7-8} \cmidrule(r){9-10}
        & & dc:ti & dc:desc & dc:ti & dc:desc & dc:ti & dc:desc & dc:ti & dc:desc \\
        \midrule
        \rowcolor[gray]{0.9} 
        T5 \citep{guo_longt5_2021} & Fine-tuned  & - & 0.502 & - & 0.248 & - & 0.467 & - & - \\
        \rowcolor[gray]{0.9} 
        GPT-2 \citep{mishra_automatic_2021} & Zero-shot  & 0.123 & - & 0.044 & - & 0.170 & - & - & - \\
        \rowcolor[gray]{0.9} 
        Custom \citep{mishra_automatic_2021} & Custom  & 0.340 & - & 0.156 & - & 0.308 & - & - & - \\
        Llama 3.1 8B & Zero-shot & 0.406 & 0.449 & 0.192 & {\fontsize{9}{9}\selectfont\textbf{0.195}} & 0.329 & 0.259 & 0.661 & 0.873 \\
        Llama 3.1 8B & Few-shot & 0.414 & 0.371 & 0.201 & 0.169 & 0.339 & 0.219 & 0.661 & 0.862 \\
        Llama 3.1 70B & Zero-shot & 0.404  & 0.428 & 0.187 & 0.167 & 0.323 & 0.244 & 0.664 & 0.853 \\
        Llama 3.1 70B & Few-shot & 0.406 & 0.411 & 0.192 & 0.160  & 0.329 & 0.234 & 0.661 & 0.842 \\
        Llama 3.2 3B & Zero-shot & 0.406 & 0.449 & 0.192 & 0.188 & 0.334 & 0.256 & 0.657 & 0.875 \\
        Llama 3.2 3B & Few-shot & 0.406 & 0.409 & 0.195 & 0.163 & 0.336 & 0.233 & 0.653 & 0.836 \\
        Gemini 1.5 Flash & Zero-shot & 0.416 & 0.476 & 0.197 & 0.187 & 0.341 & 0.269 & 0.667 & {\fontsize{9}{9}\selectfont\textbf{0.885}} \\
        Gemini 1.5 Flash & Few-shot & 0.431 & {\fontsize{9}{9}\selectfont\textbf{0.477}} & 0.211 & 0.187 & 0.358 & {\fontsize{9}{9}\selectfont\textbf{0.271}} & 0.668 & {\fontsize{9}{9}\selectfont\textbf{0.885}} \\
        Gemini 1.5 Pro & Zero-shot & 0.422 & 0.464 & 0.198  & 0.168 & 0.350 & 0.258 & 0.664 & 0.884 \\
        Gemini 1.5 Pro & Few-shot & {\fontsize{9}{9}\selectfont\textbf{0.441}} & 0.465 & {\fontsize{9}{9}\selectfont\textbf{0.219}} & 0.169 & {\fontsize{9}{9}\selectfont\textbf{0.372}} & 0.259 & {\fontsize{9}{9}\selectfont\textbf{0.672}} & {\fontsize{9}{9}\selectfont\textbf{0.885}} \\
        GPT-4o Mini & Zero-shot & 0.398 & 0.463 & 0.182 & 0.168 & 0.313 & 0.253 & 0.664 & 0.879 \\
        GPT-4o Mini & Few-shot & 0.411 & 0.465 & 0.193 & 0.169 & 0.336 & 0.255  & 0.664 & 0.878 \\
        GPT-4o & Zero-shot & 0.400 & 0.459 & 0.187 & 0.180 & 0.323  & 0.256 & 0.661 & 0.883 \\
        GPT-4o & Few-shot & 0.400 & 0.460 & 0.195 & 0.182 & 0.344 & 0.258 & 0.664 & 0.881 \\
        \bottomrule
    \end{tabular}
    \caption{Results: dcterms:title (dc:ti) and dcterms:description (dc:des).}
    \label{table:titleanddescription}
\end{table*}

\textbf{ROUGE} evaluates the quality of automatically generated text by comparing it with a reference sequence. This approach is based on various forms of n-gram and subsequence overlap between the candidate and reference texts, assessing the similarity in terms of lexical and structural alignment.
ROUGE-1 evaluates the single word overlap between a candidate summary and the reference summaries. ROUGE-2 examines two-word overlap, capturing short phrase similarities between the candidate and reference summaries. ROUGE-L measures the longest common sub sequence to assess similarity \citep{lin_rouge_2004}. ROUGE was applied to dcterms:title and dcterms:description. We chose ROUGE over BLEU and other frameworks for its superior recall and content overlap capture, making it more suitable for summarization tasks.

\textbf{Cosine Similarity} determines the cosine of the angle formed between two vectors. This angle-based approach is useful in comparing the directionality of vectors, making it a popular measure in applications like evaluating semantic similarity between high-dimensional objects (e.g., word embeddings). By normalizing the vectors, cosine similarity produces values between -1 and 1, where higher values indicate a greater similarity \citep{steck_is_2024}. This metric was applied to dcterms:title, dcterms:description, dcat:keyword and dcat:theme. We chose cosine similarity as a complementary metric to capture semantic meaning beyond lexical matches in micro F1 or ROUGE.

\textbf{Fuzzy String Matching} is a technique used to find close matches for a given sequence when an exact match does not exist. One common fuzzy matching algorithm, which was utilized in this work, is the Levenshtein distance, which calculates the ``edit distance'' between two strings. This distance represents the smallest number of single-character modifications needed to convert one string into another. A shorter distance indicates greater similarity between the two strings \citep{akila_fuzzy_2019}. This metric was applied to dcterms:creator to handle variations and inconsistencies in names, where exact matches may not always be necessary.

%% file: sections/05_discussion.tex
\section{Results}
\label{sec:discussion}

Tables 1–5 present the results. 
The best results for each metric are highlighted in bold, while benchmark results (when available) are underlined in gray.

Analysis of performance in text generation tasks for \textbf{dcterms:title} and \textbf{dcterms:description} revealed larger differences between model families than within them, with Gemini models slightly outperforming Llama and GPT, see Table \ref{table:titleanddescription}. 
Although all of the tested models achieved strong ROUGE-1 scores—either surpassing or closely matching benchmarks and indicating robust lexical matching—their lower ROUGE-2 and moderate ROUGE-L scores highlighted challenges with structural coherence for both DCAT properties.

However, all tested models achieved moderate cosine similarity values for dcterms:title (0.653–0.672) and high values for dcterms:description (0.836–0.885), demonstrating their ability to capture the conceptual essence even if they fall short in achieving the nuanced phrasing, consistent style, and flow of human-generated text. Few-shot prompting provided only incremental improvements, with a notable distinction between model sizes: larger models showed slight enhancements with additional context, whereas smaller models exhibited no improvement or even degraded performance, particularly for dcterms:description.

\begin{table}[b]
    \centering
    \begin{tabular}{@{}llcc@{}}
        \toprule
        \textbf{Model} & \textbf{Prompting} & \multicolumn{2}{c}{\textbf{Precision}} \\
        \cmidrule(r){3-4}
        & & 100\%  & 90\% \\
        \midrule
        Llama 3.1 8B & Zero-shot & 0.671 & 0.709  \\
        Llama 3.1 8B & Few-shot & 0.394 & 0.417 \\
        Llama 3.1 70B & Zero-shot & 0.888 & 0.926  \\
        Llama 3.1 70B & Few-shot & 0.881 & 0.922\\
        Llama 3.2 3B & Zero-shot & 0.834 & 0.892 \\
        Llama 3.2 3B & Few-shot & 0.840 & 0.890 \\
        Gemini 1.5 Flash & Zero-shot & 0.832 & 0.880 \\
        Gemini 1.5 Flash & Few-shot & 0.847 & 0.895 \\
        Gemini 1.5 Pro & Zero-shot & 0.877 & 0.926 \\
        Gemini 1.5 Pro & Few-shot & 0.886 & 0.929 \\
        GPT-4o Mini & Zero-shot & 0.892 & 0.923 \\
        GPT-4o Mini & Few-shot & 0.895 & 0.928 \\
        GPT-4o & Zero-shot & 0.906 & 0.936 \\
        GPT-4o & Few-shot & {\fontsize{10}{10}\selectfont\textbf{0.907}} & {\fontsize{10}{10}\selectfont\textbf{0.936}} \\
        \bottomrule
    \end{tabular}
    \caption{Results: dcterms:creator with exact match (100\%) and 90\% fuzzy matching threshold.}
    \label{table:creator}
\end{table}

\begin{table*}[t]
    \centering
    \begin{tabular}{@{}llcccc@{}}
        \toprule
        \textbf{Model} & \textbf{Prompting} & \multicolumn{4}{c}{\textbf{Precision}} \\
        \cmidrule(r){3-6}
        & & \multicolumn{2}{c}{dcterms:language} & dcterms:spatial & dcat:issued \\
        \cmidrule(r){3-4} \cmidrule(r){5-5} \cmidrule(r){6-6}
        & & Top 10 languages + GER & Rest & & \\
        \midrule
        Llama 3.1 8B & Zero-shot & 0.691 & 0.573 & 0.977 & 0.994 \\
        Llama 3.1 8B & Few-shot & 0.818  & 0.576 & 0.980& 0.997 \\
        Llama 3.1 70B & Zero-shot & {\fontsize{10}{10}\selectfont\textbf{1}} & 0.869 & 0.984 & 0.983 \\
        Llama 3.1 70B & Few-shot & {\fontsize{10}{10}\selectfont\textbf{1}} & 0.922 & {\fontsize{10}{10}\selectfont\textbf{0.987}} & 0.997 \\
        Llama 3.2 3B & Zero-shot & 0.982 & 0.460 & 0.971 & 0.300 \\
        Llama 3.2 3B & Few-shot & 0.564 & 0.327 & 0.984 & 0.968 \\
        Gemini 1.5 Flash & Zero-shot & {\fontsize{10}{10}\selectfont\textbf{1}} & 0.847 & 0.984 & 0.984 \\
        Gemini 1.5 Flash & Few-shot & {\fontsize{10}{10}\selectfont\textbf{1}}  & 0.858 & 0.980 & 0.998 \\
        Gemini 1.5 Pro & Zero-shot & {\fontsize{10}{10}\selectfont\textbf{1}}  & 0.880 & 0.984 & 0.963 \\
        Gemini 1.5 Pro & Few-shot & {\fontsize{10}{10}\selectfont\textbf{1}}  & 0.858 & 0.977 & 0.992 \\
        GPT-4o Mini & Zero-shot & {\fontsize{10}{10}\selectfont\textbf{1}}  & 0.880 & 0.984 & 0.970 \\
        GPT-4o Mini & Few-shot & {\fontsize{10}{10}\selectfont\textbf{1}}  & 0.907 & {\fontsize{10}{10}\selectfont\textbf{0.987}} & 0.997 \\
        GPT-4o & Zero-shot & {\fontsize{10}{10}\selectfont\textbf{1}}  & 0.903 & - & 0.990 \\
        GPT-4o & Few-shot & {\fontsize{10}{10}\selectfont\textbf{1}}  & {\fontsize{10}{10}\selectfont\textbf{0.936}} & - & {\fontsize{10}{10}\selectfont\textbf{0.999}} \\
        \bottomrule
    \end{tabular}
    \caption{Results: dcterms:language,  dcterms:spatial and dcat:issued.}
    \label{table:languagelocationissued}
\end{table*}

For \textbf{dcterms:creator}, all models except Llama 3.1 8B achieved precision above 0.880 for the 90\% threshold for author recognition, see Table \ref{table:creator}. The differences between the remaining models and prompting techniques were minimal. However, as the threshold increased to 100\%, larger models outperformed smaller ones, with GPT-4o achieving the highest precision score of 0.907. This suggests that models with higher parameter counts are better equipped to deliver greater accuracy.

Considering \textbf{dcterms:language}, most models achieved perfect scores (1.0) for the top 10 languages and German, see Table \ref{table:languagelocationissued}. GPT models excelled across all languages, with GPT-4o achieving a precision of 0.936 even for rarer languages. By contrast, the smaller Llama models struggled with language recognition—particularly for rarer languages, likely due to limited training data and weaker generalization capabilities. However, the strong performance of Llama 3.1 70B shows that increasing model parameters can significantly enhance multilingual understanding, even though it, like its smaller counterparts, officially supports only eight languages.

For \textbf{dcterms:spatial} and \textbf{dcat:issued}, most models achieved high precision scores, approaching 1, see Table \ref{table:languagelocationissued}. Few-shot prompting offered only a marginal advantage over zero-shot prompting for these two tasks. The only outlier here was Llama 3.2 3B for date detection which showed a weak performance in zero-shot prompting but was able to close this performance gap to the other models in few-shot prompting. These results suggest that the computational requirements for this task are relatively moderate.

\begin{table*}[t]
    \centering
    \begin{tabular}{@{}llccccccccc@{}}
        \toprule
        \textbf{Model} & \textbf{Prompting} & \multicolumn{2}{c}{\textbf{F1@5}} & \multicolumn{2}{c}{\textbf{F1@10}} & \multicolumn{2}{c}{\textbf{F1@15}} & \multicolumn{2}{c}{\textbf{Cosine-Sim.}} \\
        \cmidrule(r){3-4} \cmidrule(r){5-6} \cmidrule(r){7-8} \cmidrule(r){9-10}
        & & SE10 & SE17 & SE10 & SE17 & SE10 & SE17 & SE10 & SE17 \\
        \midrule
        \rowcolor[gray]{0.9} 
        T5 \citep{kong_promptrank_2023} & Custom & 0.172 & 0.271 & 0.207 & 0.378 & 0.214 & 0.416 & - & - \\
        Llama 3.1 8B & Zero-shot & 0.193 & 0.239 & 0.193 & 0.311 & 0.162 & 0.294 & 0.766 & 0.884 \\
        Llama 3.1 8B & Few-shot & 0.138 & 0.245 & 0.151 & 0.329 & 0.129 & 0.306 & 0.764 & 0.888 \\
        Llama 3.1 70B & Zero-shot & 0.207 & 0.235 & 0.228 & 0.308 & 0.207 & 0.306 & 0.820 & 0.904 \\
        Llama 3.1 70B & Few-shot & 0.211 & 0.225 & 0.235 & 0.299 & 0.218 & 0.292 & 0.840 & 0.901 \\
        Llama 3.2 3B & Zero-shot & 0.206 & {\fontsize{10}{10}\selectfont\textbf{0.256}} & 0.207 & 0.293 & 0.175 & 0.248 & 0.802 & 0.854 \\
        Llama 3.2 3B & Few-shot & 0.211 & 0.243 & 0.221 & 0.320 & 0.190 & 0.290 & 0.811 & 0.877 \\
        Gemini 1.5 Flash & Zero-shot & {\fontsize{10}{10}\selectfont\textbf{0.226}} & 0.233 & {\fontsize{10}{10}\selectfont\textbf{0.341}} & 0.322 & 0.243 & 0.349 & 0.836 & 0.907 \\
        Gemini 1.5 Flash & Few-shot & 0.225 & 0.241 & 0.264 & 0.341 & {\fontsize{10}{10}\selectfont\textbf{0.266}} & 0.374 & 0.842 & 0.910\\
        Gemini 1.5 Pro & Zero-shot & 0.198 & 0.212 & 0.225 & 0.296 & 0.217 & 0.310 & 0.836 & 0.905 \\
        Gemini 1.5 Pro & Few-shot & 0.219 & 0.219 & 0.252 & 0.297 & 0.240 & 0.289 & 0.834 & 0.892 \\
        GPT-4o Mini & Zero-shot & 0.184 & 0.227 & 0.200 & 0.309 & 0.187 & 0.325 & 0.832 & 0.907 \\
        GPT-4o Mini & Few-shot & 0.202 & 0.235 & 0.233 & 0.323 & 0.221 & 0.342 & {\fontsize{10}{10}\selectfont\textbf{0.855}} & 0.912 \\
        GPT-4o & Zero-shot & 0.176 & 0.231 & 0.215 & 0.330 & 0.220 & 0.366 & 0.833 & {\fontsize{10}{10}\selectfont\textbf{0.914}} \\
        GPT-4o & Few-shot & 0.197 & 0.244 & 0.236 & {\fontsize{10}{10}\selectfont\textbf{0.349}} & 0.248 & {\fontsize{10}{10}\selectfont\textbf{0.391}} & 0.843 & {\fontsize{10}{10}\selectfont\textbf{0.914}}\\
        \bottomrule
    \end{tabular}
    \caption{Results: dcat:keyword SemEval2010 (SE10) and SemEval2017 (SE17).}
    \label{table:keyword}
\end{table*}

For \textbf{dcat:keyword}, the results are displayed in Table \ref{table:keyword}. Llama 3.2 3B presents an interesting case: despite being the smallest model in the comparison, it performs exceptionally well on F1@5 scores for both SemEval datasets, outperforming larger models like GPT-4o. However, this strong initial performance comes with a notable limitation: its accuracy declines significantly at higher F1 metrics, revealing a struggle to sustain performance as task complexity increases. While few-shot prompting provides some improvement, it fails to fully address this limitation.

The analysis highlights a clear link between model size, task complexity, and performance. Larger models consistently outperform smaller ones as tasks demand more labels, maintaining better accuracy while smaller models face increasing limitations. This trend is evident in cosine similarity scores, where larger models generate keywords more closely aligned with human annotations. The performance gap becomes more pronounced with longer input sequences, demonstrating that higher parameter counts enhance the ability to process extended inputs. Additionally, larger models benefit more significantly from few-shot prompting compared to smaller ones, while smaller models, such as Llama 3.1 8B, even exhibit a decrease in performance for certain metrics under few-shot prompting. 

Considering \textbf{dcat:theme},
the results shown in Table \ref{table:theme} demonstrate that fine-tuned models are still better at matching document content with an appropriate label compared to foundational models. Fine-tuning adapts model parameters to specific domains and classification schemes, connecting general language understanding with the task's semantic requirements. This allows models to recognize important textual patterns, word choices, and context that would be missed without domain-specific training. If fine-tuning isn't possible due to data limitations or other constraints, larger foundational models assign labels more accurately than smaller ones. The performance gap widens with more labels available for assignment.

\begin{table}[b]
\centering
\resizebox{\columnwidth-4mm}{!}{%
\begin{tabular}{@{}llccc@{}}
    \toprule
    \textbf{Model} & \textbf{Prompting} & \multicolumn{2}{c}{\textbf{Micro F1}} & \textbf{Cos-Sim} \\
    \cmidrule(r){3-4} \cmidrule(r){5-5}
    & & 57K & GIZ & 57K \\
    \midrule
    \rowcolor[gray]{0.9} 
    BERT \citep{chalkidis_large-scale_2019} & Fine-tuned & 0.732 & - & - \\
    \rowcolor[gray]{0.9}
    BERT \citep{sun_laficmil_2024} & Custom & 0.737 & - & - \\
    \rowcolor[gray]{0.9}
    BERT \citep{singh_mpnet-multilabel-sector-classifier_2023} & Fine-tuned & - & 0.846 & - \\
    Gemini 1.5 Flash & Fine-tuned & {\fontsize{10}{10}\selectfont\textbf{0.547}} & {\fontsize{10}{10}\selectfont\textbf{0.794}} & {\fontsize{10}{10}\selectfont\textbf{0.833}} \\
    Llama 3.1 8B & Zero-shot & 0.257 & 0.540 & 0.766 \\
    Llama 3.1 8B & Few-shot & 0.330 & 0.632 & 0.790 \\
    Llama 3.1 70B & Zero-shot & 0.271 & 0.587 & 0.753 \\
    Llama 3.1 70B & Few-shot & 0.279 & 0.673 & 0.752\\
    Llama 3.2 3B & Zero-shot & 0.208 & 0.433 & 0.726 \\
    Llama 3.2 3B & Few-shot & 0.341 & 0.589 & 0.780 \\
    Gemini 1.5 Flash & Zero-shot & 0.268 & 0.575 & 0.760 \\
    Gemini 1.5 Flash & Few-shot & 0.345 & 0.666 & 0.780 \\
    Gemini 1.5 Pro & Zero-shot & 0.295 & 0.584 & 0.764 \\
    Gemini 1.5 Pro & Few-shot & 0.422 & 0.679 & 0.805 \\
    GPT-4o Mini & Zero-shot & 0.251 & 0.583 & 0.746 \\
    GPT-4o Mini & Few-shot & 0.288 & 0.669 & 0.757 \\
    GPT-4o & Zero-shot & 0.253 & 0.634 & 0.744 \\
    GPT-4o & Few-shot & 0.434 & 0.694 & 0.816 \\
    \bottomrule
\end{tabular}
}
\caption{Results: dcat:theme EUROLEX57K (57K) and GIZ Sector Data (GIZ).}
\label{table:theme}
\end{table}

Smaller models benefit more significantly from few-shot prompting when the labels are few and semantically distinct compared to larger models, thereby narrowing the performance gap with larger models. For instance, Llama 3.2 3B demonstrates an over 36\% gain in micro F1 on the GIZ dataset, compared to only about 9\% for GPT-4o. Conversely, when labels are numerous and semantically similar, both smaller and larger models show significant performance improvements when provided with examples. On EUROLEX57K, GPT-4o achieves over a 70\% gain in micro F1, while Llama 3.2 3B shows a 63\% increase.

Although neither foundational nor fine-tuned decoder-only models achieve the precision of a fine-tuned encoder-only model like BERT, all tested LLMs still assign labels that closely align semantically with human annotations. This alignment is reflected in high cosine similarity scores—ranging from 0.726 for Llama 3.2.3B in zero-shot mode to 0.833 for the fine-tuned Gemini 1.5 Flash. While the micro F1 scores of these two models differ significantly (0.208 vs. 0.547), their assigned labels demonstrate a high degree of shared semantic meaning.

%% file: sections/06_contribution.tex
\section{Discussion}
Up to now, it was not fully understood to what extent data catalogs can leverage LLMs to automatically extract metadata.
Our work provides the theoretical foundation on which the data catalog community can build to improve their practical tools and advance knowledge through future research.

Our evaluation revealed distinct performance patterns based on task complexity and model characteristics. For basic information extraction tasks, all models performed well regardless of size. However, in complex tasks requiring deep contextual understanding, larger models demonstrated clear advantages.
Text generation tasks showed that all models could effectively capture semantic meaning and domain-appropriate vocabulary, though they struggled with nuanced phrasing characteristic of human writing. Classification tasks revealed a strong correlation between model size and performance, particularly when handling semantically similar labels. Notably, fine-tuned decoder-only models consistently underperformed compared to encoder-only counterparts in classification tasks.

Few-shot prompting emerged as a key differentiator, with varying effects based on model size and task complexity. Smaller models showed performance improvements with few-shot prompting on simple tasks but experienced degradation when handling complex tasks with increased context load. Larger models effectively leveraged additional context across all scenarios, particularly excelling in complex tasks.

The study proposes a tiered approach for the utilization of LLMs in data catalogs for metadata generation.
For basic tasks like information extraction, smaller models offer cost-effective solutions suitable for large-scale catalogs or resource-constrained environments.
However, complex tasks requiring deeper context or specialized language benefit from larger models.
Furthermore, a tiered implementation strategy based on model families is proposed, leveraging each model's strengths: Gemini for coherent titles and abstracts, GPT for language identification, complex keyword generation, and classification.
Lastly, smaller Llama models for straightforward tasks such as information and feature extraction. This approach optimizes both performance and cost-effectiveness.

While LLMs demonstrate strong capabilities, human oversight remains essential, particularly in sensitive or regulated domains. Organizations should implement hybrid workflows that combine AI efficiency with expert validation to ensure accuracy and compliance. While concerns such as hallucination are significantly mitigated in structured metadata extraction tasks—owing to input constraints that anchor outputs to source content (e.g., title generation from abstracts)—prudent validation remains advisable in scenarios with ambiguous inputs. These structured tasks inherently limit the model's creative latitude, reducing hallucination risks compared to open-ended generation, where outputs lack grounding in predefined data.

The emergence of mobile-optimized models and high-performing models like Llama 3.2 3B expands accessibility, enabling real-time metadata generation in resource-limited settings. These advancements also support the development of local, private AI-driven data catalogs and other applications.
The key recommendation is maintaining flexibility in implementation - organizations should regularly evaluate their systems, considering both performance and cost-efficiency, while staying informed about emerging models that could optimize their metadata management processes.

%% file: sections/08_conclusion.tex
\section{Conclusion}
\label{sec:conclusion}
We return to the three hypotheses raised:

\textbf{H1.}
Autoregressive LLMs can generate DCAT-compatible metadata comparable to human-curated metadata in many scenarios, though their suitability varies by use case. They excel in generating semantically aligned titles and abstracts, reliably perform information extraction, and achieve moderate to high accuracy in classification and label creation while effectively capturing overall semantic meaning. Few-shot prompting improves precision and recall, benefiting smaller models on simpler tasks by bridging the gap to larger models, while enhancing complex reasoning for larger models.

\textbf{H2.}
Domain-specific fine-tuning significantly improves classification performance, especially with numerous semantically similar labels. While few-shot prompting helps provide examples, it cannot fully capture subtle contextual relationships. Fine-tuning adapts model parameters to specific domains, making them more sensitive to particular word choices and contextual hints.

\textbf{H3.}
Larger LLMs often outperform smaller ones, particularly in tasks requiring complex semantic reasoning and contextual interpretation. However, this performance gap varies by task—for simpler, constrained assignments, smaller models can match larger ones' accuracy while being more resource-efficient. The relationship between model size and performance remains task-dependent.

We recognize the following limitations. The standardized use of identical prompts across model families may have unintentionally favored specific architectures, given LLMs' sensitivity to prompt design. Furthermore, the subjectivity of human-generated metadata complicates evaluation, as ground truth often reflects personal interpretation rather than absolute correctness, making alternative valid annotations hard to classify as errors. External validity concerns focus on the generalizability of findings, as the research may have limited transferability to other domains where metadata generation dynamics differ across content types or industries. 

Future work should apply this methodology to other domains, such as news media or social platforms, to validate LLMs' metadata generation capabilities. Consistent performance would confirm robustness across industries, while variations would reveal domain-specific challenges, like specialized terminology and structural differences, warranting further investigation for broader optimization. Further research should incorporate expert validation to complement quantitative metrics with real-world qualitative assessments, ensuring alignment with domain standards and usability needs.